\documentclass[conference]{IEEEtran}
\IEEEoverridecommandlockouts
\usepackage{cite}
\usepackage{amsmath,amssymb,amsfonts}
\usepackage{algorithmic}
\usepackage{graphicx}
\usepackage{textcomp}
\usepackage{xcolor}

\usepackage{subfigure}
\usepackage{bm}
\usepackage{algorithm}
\usepackage{multirow}
\usepackage{booktabs}
\usepackage[bookmarks=false]{hyperref}

\def\BibTeX{{\rm B\kern-.05em{\sc i\kern-.025em b}\kern-.08em
    T\kern-.1667em\lower.7ex\hbox{E}\kern-.125emX}}
\begin{document}

\title{Learning Non-Unique Segmentation with Reward-Penalty Dice Loss}

\author{Jiabo He\textsuperscript{1}, Sarah Erfani\textsuperscript{1}, Sudanthi Wijewickrema\textsuperscript{2}, Stephen O’Leary\textsuperscript{2}, Kotagiri Ramamohanarao\textsuperscript{1}\\
\textsuperscript{1}School of Computing and Information Systems, The University of Melbourne, Australia\\
\textsuperscript{2}Department of Otolaryngology, The University of Melbourne, Australia\\
Email: \{jiaboh@student., sarah.erfani@, sudanthi.wijewickrema@, sjoleary@, kotagiri@\}unimelb.edu.au
}

\maketitle

\begin{abstract}
  Semantic segmentation is one of the key problems in the field of computer vision, as it enables computer image understanding. However, most research and applications of semantic segmentation focus on addressing unique segmentation problems, where there is only one gold standard segmentation result for every input image. This may not be true in some problems, e.g., medical applications. We may have non-unique segmentation annotations as different surgeons may perform successful surgeries for the same patient in slightly different ways. To comprehensively learn non-unique segmentation tasks, we propose the reward-penalty Dice loss (RPDL) function as the optimization objective for deep convolutional neural networks (DCNN). RPDL is capable of helping DCNN learn non-unique segmentation by enhancing common regions and penalizing outside ones. Experimental results show that RPDL improves the performance of DCNN models by up to 18.4\% compared with other loss functions on our collected surgical dataset.
\end{abstract}


\section{Introduction}
\begin{figure}[pt]
    \centering
    \subfigure[Original]
    {
    \includegraphics[width=0.14\textwidth]
    {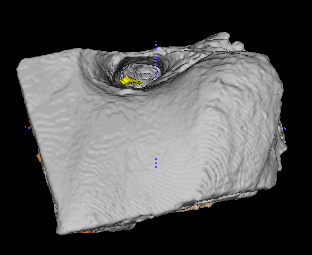}
    }
    \subfigure[Surgeon1]
    {
    \includegraphics[width=0.14\textwidth]
    {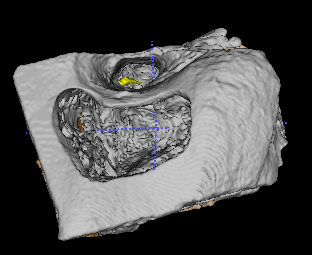}
    }
    \subfigure[Surgeon2]
    {
    \includegraphics[width=0.14\textwidth]
    {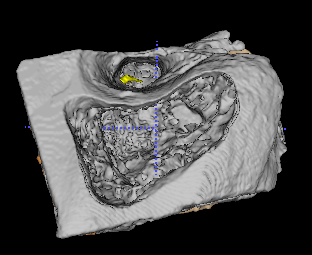}
    }
    \subfigure[Surgeon3]
    {
    \includegraphics[width=0.14\textwidth]
    {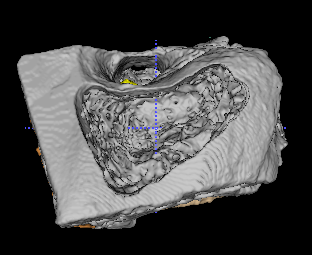}
    }
    \subfigure[Surgeon4]
    {
    \includegraphics[width=0.14\textwidth]
    {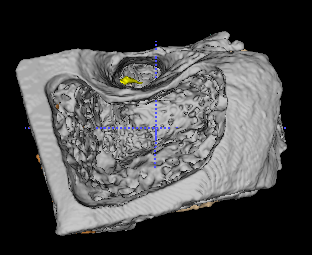}
    }
    \subfigure[Surgeon5]
    {
    \includegraphics[width=0.14\textwidth]
    {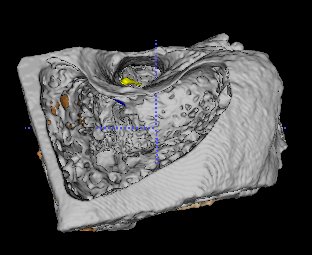}
    }
    \caption{Examples of the original temporal bone of a patient and corresponding surgical regions performed by different surgeons in the cortical mastoidectomy surgery.}
    \label{image:example}
\end{figure}

Semantic segmentation is pixel-wise image classification applied in a variety of scenarios \cite{treml2016speeding, luc2017predicting, milioto2019bonnet, cadena2016multi}. The importance of semantic segmentation is due to the fact that it can help models understand the context in the environment they are operating. Deep convolutional neural networks (DCNN) are now the de facto standard in semantic segmentation tasks because of their state-of-the-art performance \cite{long2015fully, ronneberger2015u, milletari2016v}. To train DCNN models, researchers usually build the dataset by annotating the unique pixel-wise segmentation output for every input as the ground truth. However, this may not be the case in some surgical scenarios. Different surgeons may provide non-unique surgical procedures when performing the same surgery on the same patient. Generating a model from non-unique segmentation is essential for training purposes, e.g., designing surgery simulators, verifying outcomes, and learning from other surgeons. A typical example for this is cortical mastoidectomy (CM), where surgeons remove part of the mastoid bone. The CM surgery can be performed to remove diseased bones or as a preliminary step of the cochlear implant surgery, which is an effective surgery to help patients recover from hearing loss \cite{blamey2013factors, cochlear2012}. Surgeons follow a set of guidelines to achieve this (such as identifying landmarks and drilling parallel to anatomical structures), but there is also some leeway as to how the procedure is performed. As such, the end products drilled by different surgeons may not be the same even for the same patient (Fig. \ref{image:example}). 

% Trainees and surgeons usually practise surgeries in the virtual reality surgery simulator. It is significant not only for trainees to learn the CM surgery comprehensively from different expert surgeons, but also for experts to learn from each other in the simulator. 

The non-unique segmentation can also be defined as a semantic segmentation task where every input corresponds to multiple segmentation annotation outputs. DCNN models are designed for accurate pixel-wise image segmentation in semantic segmentation tasks. Fully convolutional networks (FCN) \cite{long2015fully}, U-net \cite{ronneberger2015u} and V-net \cite{milletari2016v} are designed for specific segmentation problems while autoencoders \cite{myronenko20183d, baur2018deep} and generative adversarial networks (GANs) \cite{goodfellow2014generative, son2017retinal, xue2018segan} are also applied to certain segmentation problems. However, they are all designed to address unique segmentation problems. In the non-unique segmentation task, original images are fed into DCNN as inputs while non-unique segmentation results are regarded as outputs. The generated segmentation result by DCNN is evaluated by comparing them with all outputs corresponding to the same input. 

To train DCNN models, it is crucial to design the optimization objective based on DCNN architectures and targets. Existing loss functions, such as the cross-entropy loss (CEL) \cite{long2015fully}, the weighted cross-entropy loss (WCEL) \cite{ronneberger2015u} and the Dice loss (DL) \cite{sudre2017generalised}, are utilized as optimization objectives in semantic segmentation tasks. WCEL was derived from CEL to alleviate the imbalance problem between the foreground and the background of images \cite{ronneberger2015u}. Later, DL was proposed based on measuring overlaps between predicted outputs and the ground truth, which is more robust than WCEL when the level of imbalance increases \cite{sudre2017generalised}. However, they are limited to address the unique segmentation problem and cannot enhance the commonality of non-unique segmentation outputs for the same input.

Driven by the significance of learning non-unique segmentation and limitations of existing algorithms, a question arises: \emph{How to learn non-unique segmentation better?} In this paper, we propose the reward-penalty Dice loss (RPDL) function to address this problem. Specifically, pixel-wise positive rewards are designed to enhance common segmentation regions and pixel-wise negative rewards are designed to penalize outside ones in RPDL. In order to demonstrate advantages of RPDL over other loss functions in non-unique segmentation tasks, we first collected $63$ different CM surgeries in the simulator. We invited $7$ surgeons at the Royal Victorian Eye and Ear Hospital, with each performing one CM surgery on $9$ specimens of different patients' temporal bones. Surgeons have a variety of ways to remove part of the mastoid. The final regions of the removed volume performed by different surgeons are different from each other for every patient. It is a medical image segmentation task with original bone images regarded as inputs and corresponding surgical results from different surgeons regarded as outputs. Experimental results show that RPDL can help DCNN learn CM surgeries from different surgeons comprehensively. DCNN with RPDL outperform those with other loss functions dominantly in providing surgical regions for new patients. The automated generation of CM surgery results for new patients can not only be of high efficiency in time, but also provide valid surgeries with less variance after learning all surgeries comprehensively from different surgeons. There are three main contributions in this paper:

\begin{itemize}
  \item We define non-unique segmentation as a new semantic segmentation problem, where every input image corresponds to multiple possible outputs. We extend the potential of DCNN to learn non-unique segmentation tasks. In addition, we collect the CM dataset to provide a new benchmark dataset for non-unique segmentation\footnote{The code and dataset is available at \url{https://github.com/Jacobi93/Reward-penalty-Dice-loss}.}.
  \item We propose a new loss function, RPDL, for DCNN to learn non-unique segmentation tasks comprehensively. RPDL is able to help DCNN models enhance their ability to extract the similarity among different segmentation results and get rid of touching outside regions.
  \item We validate advantages of RPDL over other loss functions on the CM dataset that we collect. RPDL outperforms other loss functions evaluated by both evaluation metrics and expert surgeons.
\end{itemize}

\section{Related Work}

\subsection{Medical Image Segmentation}
The success of pixel-wise segmentation is attributed to the fast and effective in-network upsampling, which learns dense prediction by deconvolution in deep convolutional neural networks (DCNN) \cite{shan2008fast, long2015fully}. There have been a lot of medical image segmentation competitions and challenges owing to the development of DCNN in recent years, such as BraTS Challenge \cite{brats2019}, LiTS Challenge \cite{LiTS2017}, Atrial Segmentation Challenge \cite{atrial2018}, and Medical Segmentation Decathlon \cite{msd2018}, etc. These competitions and challenges are so competitive that the grades of top teams are very close to each other. For example, the difference among top $5$ teams are within $3\%$ in $7$ out of $10$ individual disciplines in the Medical Segmentation Decathlon 2018 Challenge, measured by the Dice coefficient \cite{msd2018}. Most top teams tend to build their models based on different architectures of DCNN. DCNN models with delicate design of architectures and fine tuning of hyperparameters share comparable and close performance in medical image segmentation tasks. 

There are a plenty of architectures of DCNN designed for medical image segmentation, among which U-net \cite{isensee2017brain, isensee2018no, isensee2018nnu}, autoencoders \cite{myronenko20183d, baur2018deep, huang2019mala}, and generative adversarial networks (GANs) \cite{goodfellow2014generative, son2017retinal, xue2018segan} are popular ones. 3D U-net \cite{cciccek20163d} and V-net \cite{milletari2016v} are both inspired from the U-net architecture. The major difference is that there are ResNet layers \cite{he2016deep} applied in every module of V-net instead of 3D U-net for residual learning. U-net is actually one variation of convolutional autoencoders with the concatenation of outputs from the encoder and inputs from the decoder at the same depth. Other autoencoders, such as dense autoencoders, spatial autoencoders, and variational autoencoders may all obtain competitive performance in medical image segmentation \cite{baur2018deep}. Unlike U-net and autoencoders, GANs are trained in a different way. GANs fool the discriminator to distinguish gold standard images from synthetic ones by generating segmentation outputs with the generator. In addition, there are also further research on improving the performance of DCNN in medical image segmentation by \textbf{(1)} embedding other modules into DCNN, like attention modules for enhancing spatial correlation features \cite{oktay2018attention} or \textbf{(2)} utilizing cascaded DCNN architectures for each sub-task \cite{wang2017automatic}. However, there is only limited improvement compared with basic DCNN models.

\subsection{Loss Functions for Semantic Segmentation}
The cross-entropy loss (CEL) function was commonly used as the optimization objective of DCNN in semantic segmentation tasks. The pixel-wise CEL evaluates each pixel individually and optimizes the summation of them \cite{long2015fully}. However, CEL cannot address the imbalance problem between the background and the foreground of images. Thus, the weighted cross-entropy loss (WCEL) function was proposed to handle the mild-imbalance problem \cite{ronneberger2015u, cciccek20163d, kampffmeyer2016semantic}. The weight is a hyperparameter in WCEL, which is often set as the ratio of the background and the foreground in order to enhance the minority target. Later, the Dice loss (DL) function was also proposed to address the high-imbalance problem, which is commonly used in 3D medical image segmentation tasks \cite{sudre2017generalised, isensee2017brain, wang2017automatic}. The intuition is that the Dice coefficient is the first choice as the evaluation metric for the performance of models in segmentation tasks. It is straightforward to set the evaluation metric as the optimization objective. DL is also a differentiable loss function enabling backpropagation of the gradient to the upstream of DCNN pipelines. DL is based on overlap measures and appears to be more robust than WCEL when the level of imbalance increases \cite{sudre2017generalised}. To address the non-unique segmentation problem, we propose the reward-penalty Dice loss (RPDL) derived from DL. Compared with other loss functions, RPDL can enhance common segmentation regions and penalize outside ones by the reward-penalty map.

\section{Methodology}
We consider the following problem: different segmentation results are provided by different annotation experts for the same image; there are both deviation and commonality among those results for every image; we train deep convolutional neural networks (DCNN) to learn the non-unique segmentation so that DCNN can provide a trustworthy segmentation output when a new image is given.

\subsection{Preliminaries}
Loss functions are designed as the optimization objective of models. In particular, some loss functions are proposed for addressing the image segmentation problem. In this paper, two commonly used loss functions for semantic segmentation are selected as baselines, namely the weighted cross-entropy loss (WCEL) and the Dice loss (DL) functions. These loss functions can all be applied to both binary segmentation tasks and multi-class segmentation tasks. In fact, multi-class segmentation tasks are usually regarded as multiple binary sub-tasks, with each corresponding to one channel of the one-hot encoded outputs. In brief, all loss functions are described below for binary segmentation tasks.

% maybe a better word than 'in brief'

\textbf{Weighted cross-entropy loss:} In order to alleviate the imbalance problem between the background and the foreground of images, WCEL was proposed as the loss function for DCNN.
\begin{equation}
WCEL=-\frac{1}{N}\sum_{n=1}^{N}[w y_n log(P_n) + 
(1-y_n) log(1-P_n)]
\label{wcel}
\end{equation}

where $N$ is the number of pixels (2D tasks) or voxels (3D tasks) in the mini-batch. $\bm{y}$ ($y_n \in \bm{y}$) is the ground truth of segmentation. $\bm{P}$ ($P_n \in \bm{P}$) is the output of DCNN, which can be obtained using the sigmoid activation function for the last layer. $P_n$ is usually clipped into $[\epsilon, 1-\epsilon]$ to avoid $\log(0)$ for training stability, where $\epsilon$ is a small default term. $w$ is the weight to alleviate the imbalance problem, which is often set as the ratio of the background and the foreground volume. For example, if the ratio of the background and the foreground in volume is $49:1$ (only $2\%$ of the total region should be segmented in average), then $w=49$ is set to the weight in WCEL.

\textbf{Dice loss:} Although WCEL can address the imbalance problem to certain extent, researchers found that DL outperforms WCEL in many segmentation tasks. The reason is that DCNN with the DL function can be optimized directly for the evaluation metric, which is the Dice coefficient. It is more robust than WCEL when the level of imbalance increases.
\begin{equation}
DL=1-\frac{2 \sum_{n=1}^{N}y_n P_n + \epsilon}
{\sum_{n=1}^{N}y_n + \sum_{n=1}^{N}P_n + \epsilon}
\label{dice}
\end{equation}

where $\epsilon$ is added to avoid the denominator to be $0$ for training stability. $\sum_{n=1}^{N}y_n P_n$ is the overlap of the ground truth $\bm{y}$ and the output $\bm{P}$.

\begin{figure}[tbp]
\centering
\includegraphics[width=9cm]{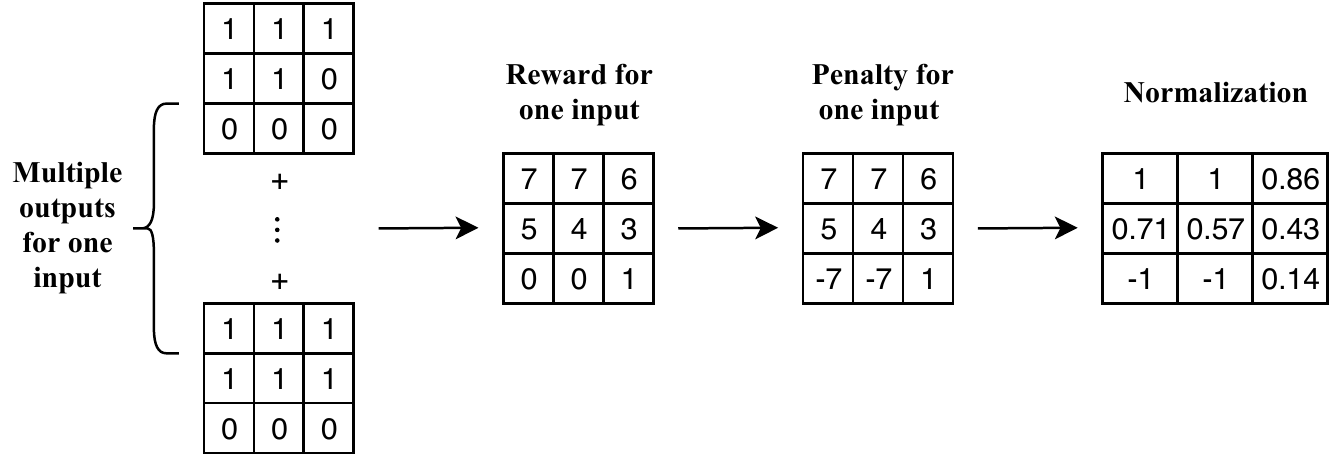}
\caption{Workflow to generate the reward-penalty map for one input.}
\label{rpmap}
\end{figure}

\subsection{Reward-penalty Dice loss}
We are finally in the position to introduce the core contribution of our work. This paper proposes the reward-penalty Dice loss (RPDL) as the derivative of DL to address the non-unique segmentation problem. When there are multiple segmentation annotations for every input image, different regions of the segmentation annotation may vary in importance. Therefore, it is crucial to extract the importance of every segmentation annotation, which is exhibited by the reward-penalty map (RPMap) $\bm{M}$. The RPDL function is inspired by enhancing common regions segmented by different experts and penalizing outside ones.
\begin{equation}
RPDL=1-\frac{2 \sum_{n=1}^{N}y_n P_n M_n + \epsilon}
{\sum_{n=1}^{N}y_n |M_n| + \sum_{n=1}^{N}P_n |M_n| + \epsilon}
\label{RPDL}
\end{equation}

where $\bm{M}$ ($M_n \in \bm{M}$) is the RPMap constructed by all possible outputs for one input in the training set. Different inputs have different corresponding RPMaps. Here is a 2D example of the procedure to generate the RPMap for one input in Fig. \ref{rpmap}. First, we construct the pixel-wise RPMap by recording the segmentation times for every pixel of the input by all experts. Second, we set the penalty to those regions that are not segmented by any expert, which is the negative of the maximum. At last, we normalize the RPMap for training. Normalization of the RPMap is empirically proven to be helpful for training stability. The penalty is a hyperparameter and we find that $-1$ after normalization is a good choice by experiments. It indicates that those regions that are not segmented by any expert should be penalized heavily if they are segmented by the model. Instead, regions segmented by different numbers of experts should be rewarded to different extent. $|M_n|$ is the absolute value of $M_n$. When a model provides a region which should be penalized, RPDL tends to be large in the model. In this case, $|M_n|$ is designed in the denominator to avoid the opposite result. Since $\bm{y}$, $\bm{P}$ and $\bm{M}$ are all in the same size, all loss functions compared in this paper can be calculated efficiently by the Hadamard product.

RPDL is also a differentiable loss function derived from DL, which is able to backpropagate its gradient to the upstream of DCNN pipelines. The gradient of RPDL with respect to the $i^{th}$ voxel of the output is obtained by

\begin{equation}
\scriptsize
\frac{\partial RPDL}{\partial P_i}=2\frac{|H_i|\sum_{n=1}^{N}y_n P_n H_n - y_i H_i (\sum_{n=1}^{N}y_n |H_n| + \sum_{n=1}^{N}P_n |H_n|)}
{(\sum_{n=1}^{N}y_n |H_n| + \sum_{n=1}^{N}P_n |H_n|)^2}
\label{RPDLg}
\end{equation}

where $\epsilon$ is not included in the gradient.

\section{The Cortical Mastoidectomy Surgery Dataset}
\begin{figure}[bp]
  \centering
  \includegraphics[width=4.5cm]{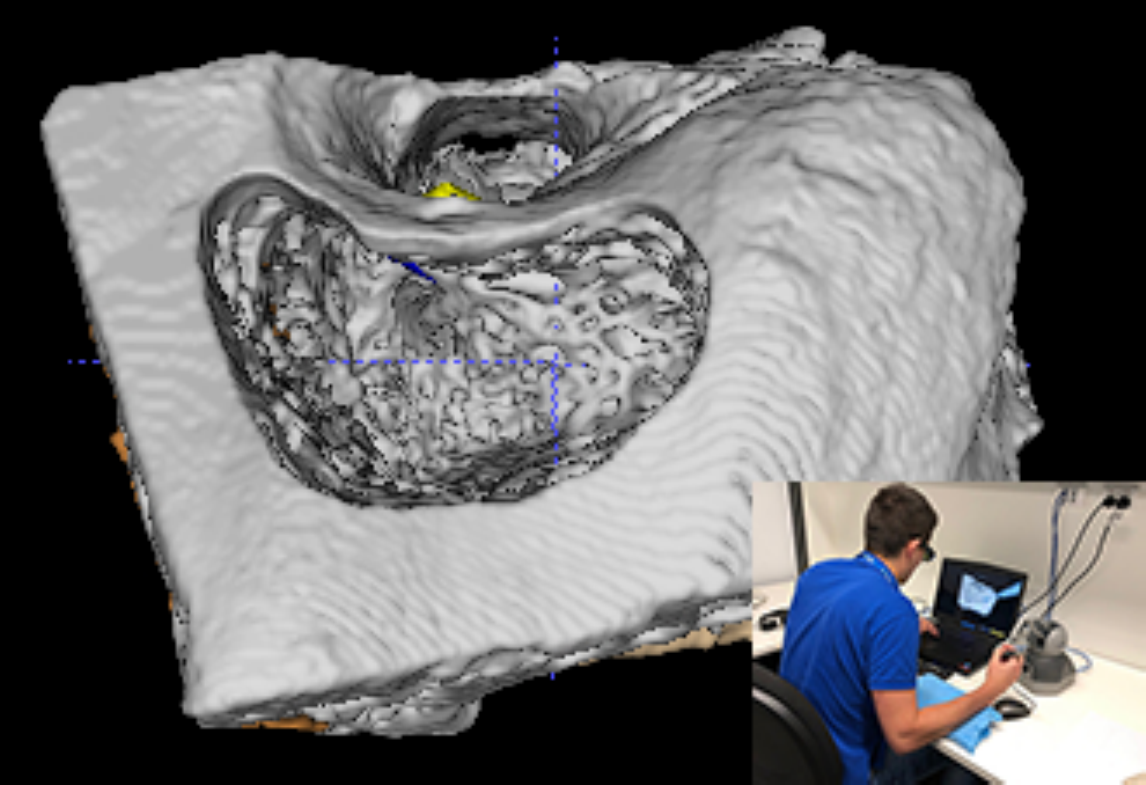}
  \caption{Temporal bone after the cortical mastoidectomy (CM) surgery (bottom right is one surgeon performing a CM surgery in the VRTBS simulator).}
  \label{temporal}
\end{figure}

In the cortical mastoidectomy (CM) surgery, the surgeon removes part of the mastoid bone to identify the incus and the facial nerve as well as avoid touching other anatomical structures such as the sigmoid sinus and the dura. Fig. \ref{temporal} is an example of a processed CT-scan temporal bone after the CM surgery in the Virtual Reality Temporal Bone Surgery (VRTBS) simulator \cite{py06nimg}. The vacant region in the center is the drilled part of the mastoid on a temporal bone. The VRTBS simulator is developed as a platform for the temporal bone surgery training, including the CM surgery. Expert surgeons can record their surgeries in the simulator so that trainees and other experts can learn from them. Trainees can also practise performing surgeries repetitively in the simulator before they achieve expertise, which is effective to minimize potential risks for patients \cite{o2008validation}. The removal of the mastoid in the simulator is consistent with effective operations of the surgery. The original CT-scan image of the temporal bone is manually segmented by expert surgeons according to different anatomical structures (categories from $0$ to $13$ represent different structures such as the air, the bone, the incus, the facial nerve, the sigmoid sinus, and the dura, etc.). All 3D images are saved with different structures annotated in different colors.

In our research, we invite $7$ surgeons at the hospital, with each performing one CM surgery on $9$ processed specimens of different patients' temporal bones, respectively. In the end, we collect $63$ different CM surgeries in the VRTBS simulator. Those processed images before surgeries are fed into DCNN as inputs and those binary segmentation results generated after surgeries are regarded as outputs. All bone images along with their segmentation outputs are various in size and we build the CM dataset by resizing each of them into $64 \times 64 \times 64$ voxels.

\section{Experimental Results and Analysis}
\begin{table*}[t]
\caption{Experimental results tested on different bones, Experiment 1 part 1 ($Dice \pm std$).}
\begin{center}
\begin{tabular}{c|c|c|c|c|c|c}
\toprule
Model&\multicolumn{3}{c|}{U-net \cite{cciccek20163d}} & \multicolumn{3}{c}{V-net \cite{milletari2016v}}\\
\midrule
Training loss & WCEL & DL & \textbf{RPDL} & WCEL & DL & \textbf{RPDL} \\
\midrule
Bone1 & $0.480 \pm 0.075$ & $0.628 \pm 0.052$ & $\textbf{0.696} \pm 0.059$ & $0.425 \pm 0.046$ & $0.584 \pm 0.030$ & $\textbf{0.625} \pm 0.063$ \\

Bone2 & $0.597 \pm 0.100$ & $0.523 \pm 0.125$ & $\textbf{0.620} \pm 0.094$ & $0.464 \pm 0.088$ & $0.414 \pm 0.103$ & $\textbf{0.538} \pm 0.098$ \\

Bone3 & $0.482 \pm 0.082$ & $0.538 \pm 0.055$ & $\textbf{0.684} \pm 0.076$ & $0.439 \pm 0.074$ & $0.458 \pm 0.060$ & $\textbf{0.506} \pm 0.065$ \\

Bone4 & $0.658 \pm 0.080$ & $0.609 \pm 0.068$ & $\textbf{0.660} \pm 0.070$ & $0.614 \pm 0.076$ & $0.703 \pm 0.052$ & $\textbf{0.725} \pm 0.072$ \\

Bone5 & $0.620 \pm 0.100$ & $0.628 \pm 0.089$ & $\textbf{0.683} \pm 0.102$ & $0.488 \pm 0.097$ & $0.322 \pm 0.071$ & $\textbf{0.532} \pm 0.107$ \\

Bone6 & $0.553 \pm 0.042$ & $0.613 \pm 0.052$ & $\textbf{0.636} \pm 0.078$ & $0.478 \pm 0.041$ & $0.553 \pm 0.041$ & $\textbf{0.559} \pm 0.048$ \\

Bone7 & $0.561 \pm 0.108$ & $\textbf{0.683} \pm 0.075$ & $0.662 \pm 0.082$ & $0.540 \pm 0.107$ & $0.641 \pm 0.100$ & $\textbf{0.693} \pm 0.097$ \\

Bone8 & $0.514 \pm 0.108$ & $0.568 \pm 0.064$ & $\textbf{0.598} \pm 0.089$ & $0.543 \pm 0.098$ & $0.571 \pm 0.066$ & $\textbf{0.574} \pm 0.056$ \\

Bone9 & $0.544 \pm 0.094$ & $\textbf{0.639} \pm 0.107$ & $0.636 \pm 0.097$ & $0.501 \pm 0.085$ & $0.598 \pm 0.066$ & $\textbf{0.600} \pm 0.056$ \\

Overall & $0.556 \pm 0.088$ & $0.603 \pm 0.076$ & $\textbf{0.653} \pm 0.083$ & $0.499 \pm 0.079$ & $0.538 \pm 0.065$ & $\textbf{0.595} \pm 0.074$ \\
\bottomrule
\end{tabular}
\label{table1}
\end{center}
\end{table*}

\begin{table*}[t]
\caption{Experimental results tested on different bones, Experiment 1 part 2 ($Dice \pm std$).}
\begin{center}
\begin{tabular}{c|c|c|c|c|c|c}
\toprule
Model &\multicolumn{3}{c|}{Isensee17 \cite{isensee2017brain}} & \multicolumn{3}{c}{Myronenko18 \cite{myronenko20183d}}\\
\midrule
Training loss & WCEL & DL & \textbf{RPDL} & WCEL & DL & \textbf{RPDL} \\
\midrule
Bone1 & $0.460 \pm 0.059$ & $0.630 \pm 0.033$ & $\textbf{0.668} \pm 0.060$ & $0.442 \pm 0.067$ & $0.661 \pm 0.056$ & $\textbf{0.677} \pm 0.066$ \\

Bone2 & $0.383 \pm 0.098$ & $\textbf{0.428} \pm 0.083$ & $0.403 \pm 0.088$ & $0.463 \pm 0.077$ & $0.676 \pm 0.086$ & $\textbf{0.680} \pm 0.119$ \\

Bone3 & $0.419 \pm 0.063$ & $0.500 \pm 0.054$ & $\textbf{0.547} \pm 0.074$ & $0.658 \pm 0.114$ & $0.624 \pm 0.054$ & $\textbf{0.665} \pm 0.072$ \\

Bone4 & $0.624 \pm 0.083$ & $0.645 \pm 0.053$ & $\textbf{0.688} \pm 0.065$ & $0.650 \pm 0.086$ & $\textbf{0.722} \pm 0.037$ & $0.692 \pm 0.056$ \\

Bone5 & $0.406 \pm 0.109$ & $0.516 \pm 0.111$ & $\textbf{0.567} \pm 0.110$ & $0.565 \pm 0.099$ & $0.689 \pm 0.085$ & $\textbf{0.697} \pm 0.108$ \\

Bone6 & $0.482 \pm 0.048$ & $0.508 \pm 0.039$ & $\textbf{0.624} \pm 0.049$ & $0.580 \pm 0.078$ & $0.644 \pm 0.042$ & $\textbf{0.680} \pm 0.063$ \\

Bone7 & $0.549 \pm 0.102$ & $0.679 \pm 0.097$ & $\textbf{0.682} \pm 0.086$ & $0.570 \pm 0.103$ & $\textbf{0.697} \pm 0.071$ & $0.677 \pm 0.076$ \\

Bone8 & $0.547 \pm 0.102$ & $0.532 \pm 0.079$ & $\textbf{0.630} \pm 0.080$ & $0.380 \pm 0.090$ & $\textbf{0.694} \pm 0.051$ & $0.680 \pm 0.099$ \\

Bone9 & $0.476 \pm 0.079$ & $0.558 \pm 0.058$ & $\textbf{0.618} \pm 0.101$ & $0.569 \pm 0.095$ & $0.585 \pm 0.091$ & $\textbf{0.676} \pm 0.093$ \\

Overall & $0.483 \pm 0.082$ & $0.555 \pm 0.067$ & $\textbf{0.603} \pm 0.079$ & $0.542 \pm 0.090$ & $0.666 \pm 0.064$ & $\textbf{0.680} \pm 0.084$ \\
\bottomrule
\end{tabular}
\label{table2}
\end{center}
\end{table*}

\textbf{Model details:} Four deep convolutional neural networks (DCNN) are selected as representatives for comparison in experiments: U-net \cite{cciccek20163d}, V-net \cite{milletari2016v}, Isensee17 \cite{isensee2017brain}, and Myronenko18 \cite{myronenko20183d}. U-net and V-net are two well-performing and widely used architectures in medical segmentation tasks. Isensee17 and Myronenko18 are two champions in BraTS Challenge 2017 and 2018, respectively. We use same architectures of all models in their original papers with minor change. For every model with different loss functions, they share the same architecture and hyperparameters. We utilize the weighted cross-entropy loss (WCEL), the Dice loss (DL) and the reward-penalty Dice loss (RPDL) as loss functions for the first 3 models, respectively. The original loss function of Myronenko18 is $L=DL + 0.1 \times L_2 + 0.1 \times L_{KL}$. We replace DL with WCEL and RPDL in comparison experiments and keep the rest terms the same.

Here are details about hyperparameters. $10\%$ of the training set is selected as the validation set in random. The size of the mini-batch is $16$ for Myronenko18 due to GPU memory limit, and $64$ for other models. The initial learning rate is $0.0001$ for all models. The decay of the learning rate is $0.1$ if the performance on the validation set does not improve in $10$ epochs, with $200$ epochs in maximum for training. The early stopping patience is $20$ epochs. The dropout rate is $0.3$ for every CNN module in the first three models. There is no dropout in Myronenko18. We use the Leaky ReLU activation after all CNN layers and the sigmoid activation after the last layer. All images in the training set are mirrored and rotated in three axes (x, y and z) so that the size of the dataset is augmented by hundreds of times, which greatly improve the performance and stability of DCNN models.

There are also some other techniques designed to improve the performance of DCNN in segmentation tasks, such as data augmentation and building an ensemble of models. However, we do not include them in our experiments because they are all orthogonal to loss functions. Data augmentation techniques, like the elastic deformation, random crops, and synthetic generation for images, are proven to be useful in most tasks, more or less \cite{bowles2018gan, frid2018gan}. They are helpful regardless of the architectures of models along with their various loss functions. Our goal is to demonstrate advantages of our proposed RPDL over other loss functions no matter what data augmentation techniques are applied on the training set. Moreover, the ensemble of models is helpful to improve the performance of DCNN to limited extent as well. It is widely applied in many competitions and challenges. Again, building an ensemble of models can improve the performance of DCNN no matter what loss functions are used in them. In conclusion, we do not apply above techniques to our experiments because they are \textbf{(1)} computation-exhausting techniques and \textbf{(2)} beyond the discussion of our paper.

\textbf{Training details:} We implement our experiments on Keras and train models on NVIDIA V100 GPUs. There are two sets of experiments done on the cortical mastoidectomy (CM) dataset in this paper. \textbf{First}, we pick each bone out along with $7$ surgeries performed on it by different surgeons as the testing set, which is the 9-fold cross validation for every model. The goal is to generate the surgical regions for new patients in the first experiment. It is very challenging for DCNN to generate comparable surgical regions to those provided by surgeons for new patients because there are only $9$ bones in the CM dataset in total, which are very different from each other. However, it is also a new benchmark with much potential for DCNN and we believe that DCNN will be able to provide surgical results competitive with those provided by surgeons, given a large size of the dataset in near future. \textbf{Second}, we pick each surgeon out along with $9$ surgeries performed on each bone. There are only 6 segmentation annotation images used for training in the second experiment, which is to evaluate if DCNN models can provide outputs more similar to segmentation annotation images than surgeons in the testing set. The second experiment is designed to compare all outputs provided by DCNN using different loss functions with surgeons in the testing set, which is not as challenging as the first experiment.

\textbf{Evaluation metrics:} The Dice coefficient is usually taken as one of the most important evaluation metrics in medical image segmentation tasks (Eq. \eqref{d}). Apart from the Dice coefficient, we also propose the reward-penalty Dice (RPD) coefficient corresponding to RPDL (Eq. \eqref{rpd}). Both evaluation metrics are shown below:

\begin{equation}
Dice=\frac{2 \sum_{n=1}^{N}y_n P'_n}
{\sum_{n=1}^{N}y_n + \sum_{n=1}^{N}P'_n}
\label{d}
\end{equation}

\begin{equation}
RPD=\frac{2 \sum_{n=1}^{N}y_n P'_n M_n}
{\sum_{n=1}^{N}y_n |M_n| + \sum_{n=1}^{N}P'_n |M_n|}
\label{rpd}
\end{equation}

\begin{figure*}
   \centering
\begin{tabular}{ccccccc}
\includegraphics[width=2.2cm]{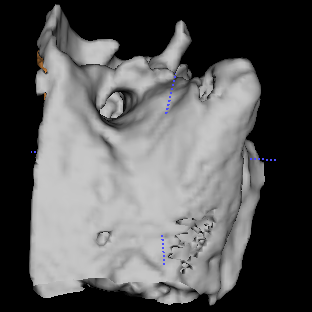}&
\includegraphics[width=2.2cm]{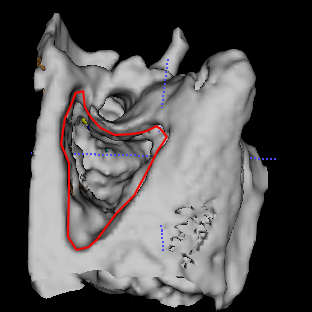}&
\includegraphics[width=2.2cm]{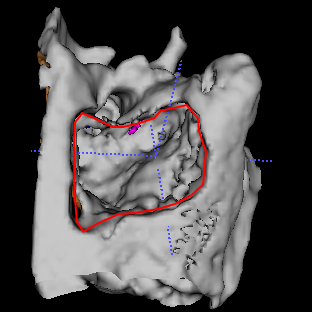}&
\includegraphics[width=2.2cm]{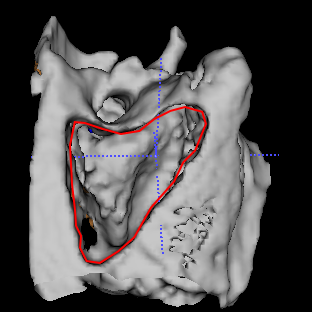}&
\includegraphics[width=2.2cm]{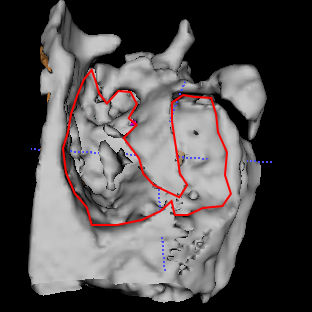}&
\includegraphics[width=2.2cm]{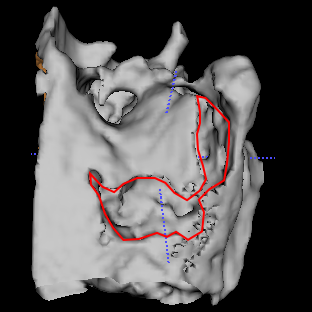}&
\includegraphics[width=2.2cm]{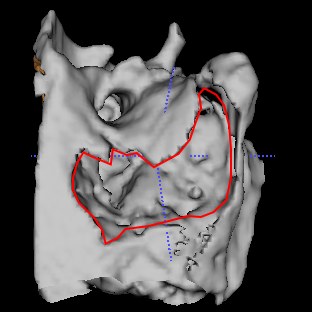}
\\

\includegraphics[width=2.2cm]{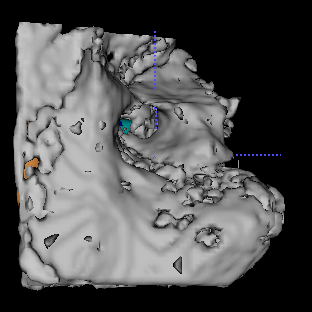}&
\includegraphics[width=2.2cm]{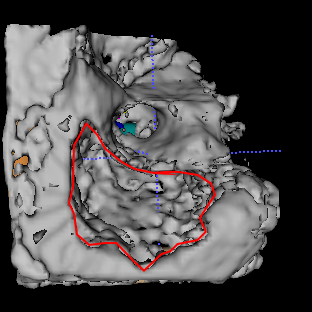}&
\includegraphics[width=2.2cm]{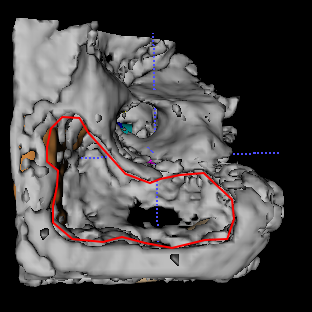}&
\includegraphics[width=2.2cm]{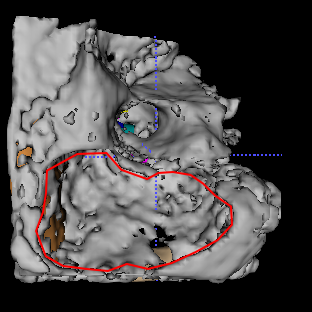}&
\includegraphics[width=2.2cm]{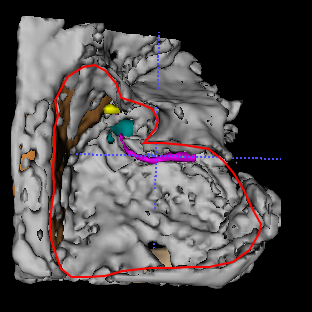}&
\includegraphics[width=2.2cm]{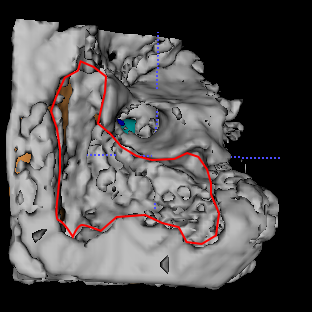}&
\includegraphics[width=2.2cm]{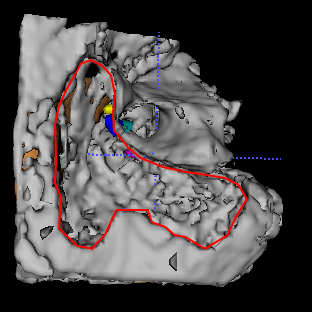}
\\

\includegraphics[width=2.2cm]{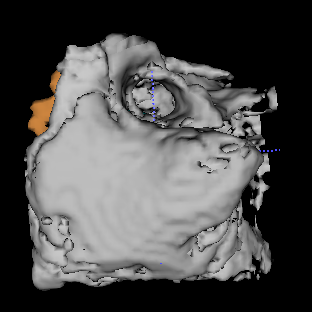}&
\includegraphics[width=2.2cm]{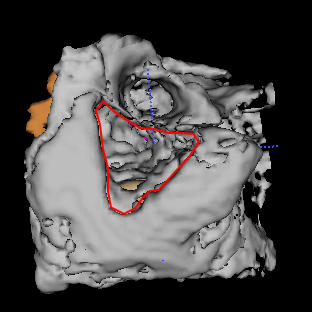}&
\includegraphics[width=2.2cm]{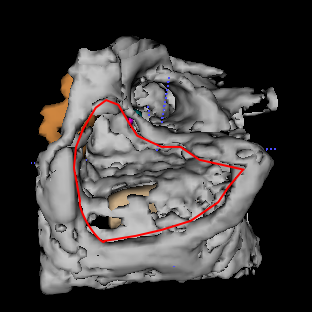}&
\includegraphics[width=2.2cm]{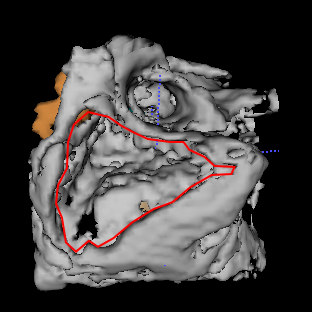}&
\includegraphics[width=2.2cm]{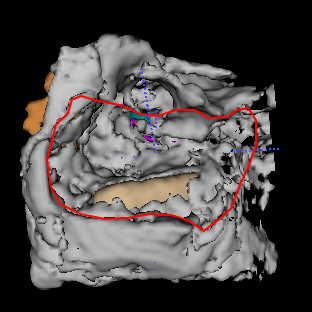}&
\includegraphics[width=2.2cm]{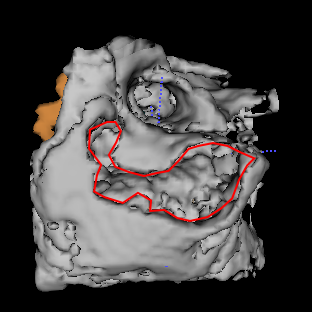}&
\includegraphics[width=2.2cm]{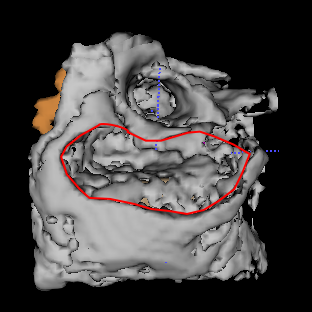}
\\

Original & Surgeon1 & Surgeon2 & Surgeon3 & WCEL & DL & RPDL
\\

\end{tabular}

    \caption{Qualitative comparisons of ground truth surgical results from surgeons and generated outputs by different loss functions in U-net in Experiment 1.}
    \label{fig:final}
\end{figure*}

where the output $\bm{P}$ after the sigmoid activation function is transformed into the binary output $\bm{P}'$, with $0.5$ as the threshold. Both Dice and RPD coefficients are large when models enjoy good performance in segmentation tasks. When testing the performance of models with different loss functions, all outputs are compared with the whole testing set to obtain overall results. For example, the segmentation output of every bone is compared with those from all surgeons in the testing set in Experiment 1. The average measurement of above evaluation metrics is then recorded as the testing performance.

\textbf{Med3D:} Med3D is a series of 3D-ResNet pre-trained models aggregating eight different 3D medical segmentation datasets \cite{chen2019med3d}. The shared encoder of Med3D can be transferred on other datasets and tasks, both accelerating the training convergence speed and improving the performance to different extent. We used the 3D-ResNet50 pre-trained Med3D as our encoder stacked by a new decoder in our experiments. However, there is no noticeable improvement of the fine-tuned stacked model's performance compared with that of representative DCNN models trained from scratch. There are two probable reasons for this issue. First, we define a new segmentation task where one input image corresponds to multiple possible outputs, which is heterogeneously transfered from all aggregated segmentation tasks in Med3D. Second, the aggregation of multi-domain datasets does not improve the performance remarkably, either. There is only about $1.96\%$ improvement in average measured by Dice after Med3D is trained on eight domains, compared with that trained in the single domain \cite{chen2019med3d}.

\subsection{Experiment 1: Training on Different Bones}

The first set of experiments picks 7 surgeries on the same bone out as the testing set at every time in the 9-fold cross validation. The goal is to make full use of all segmentation outputs in the training set and learn to do surgeries on different bones by DCNN in order to generate the most plausible surgical regions for new patients. It is a very challenging task because of the limited size of the CM dataset and the large variety of bones (Fig. \ref{fig:final} column 1). DCNN models need to know how to do the surgery on the $9^{th}$ bone based on surgeries on 8 bones from different surgeons in the training set. The large difference of the testing performance on each bone also implies the large variety of bones in the CM dataset (Table \ref{table1} and \ref{table2}).

All models with different loss functions are tested by both Dice and RPD coefficients. The overall results are consistent with each other measured by two metrics (Table \ref{table3}). The cross validation details tested by Dice are shown in Table \ref{table1} and \ref{table2}. WCEL performs the worst in all DCNN models. RPDL performs better than DL and WCEL dominantly in all models by $1.4\%-18.4\%$ (absolute), measured by both Dice and RPD on the testing set. The limited size of the training set and the large diversity of input images impede the good performance of DCNN models with all loss functions, among which RPDL performs the best. WCEL performs the worst as there is high imbalance between the background (about $96 \%$) and the foreground (about $4 \%$) of CM images. WCEL is not capable of addressing the high-imbalance problem. DL outperforms WCEL in extracting the similarity of non-unique segmentation outputs corresponding to same input images, in order to predict the surgery performed on the new bone. However, it is not enough when the training set is limited and the diversity of input images is very large. In contrast, RPDL is able to find the similarity of surgeries from different surgeons by paying more attention to common surgical regions and penalizing outside ones. In conclusion, when there are only limited training data with large diversity of input images, it is necessary to select RPDL as the loss function for DCNN. RPDL is stronger than WCEL and DL in finding the commonality among non-unique segmentation outputs.

\begin{table}[tbp]
\scriptsize
\caption{Overall results tested on different bones, Experiment 1 (up: $Dice \pm std$; down: $RPD \pm std$).}
\begin{center}
\begin{tabular}{c|c|c|c|c}
\toprule
& Training loss & WCEL & DL & \textbf{RPDL} \\
\midrule
\multirow{4}{*}{Dice} & U-net & $0.556 \pm 0.088$ & $0.603 \pm 0.076$ & $\textbf{0.653} \pm 0.083$ \\ 
& V-net & $0.499 \pm 0.079$ & $0.538 \pm 0.065$ & $\textbf{0.595} \pm 0.074$  \\
& Isensee17 & $0.483 \pm 0.082$ & $0.555 \pm 0.067$ & $\textbf{0.603} \pm 0.079$ \\
& Myronenko18 & $0.542 \pm 0.090$ & $0.666 \pm 0.064$ & $\textbf{0.680} \pm 0.084$ \\
\midrule
\multirow{4}{*}{RPD} & U-net & $0.612 \pm 0.064$ & $0.675 \pm 0.055$ & $\textbf{0.735} \pm 0.063$ \\ 
& V-net & $0.538 \pm 0.057$ & $0.604 \pm 0.043$ & $\textbf{0.661} \pm 0.052$ \\
& Isensee17 & $0.512 \pm 0.059$ & $0.623 \pm 0.044$ & $\textbf{0.671} \pm 0.062$ \\
& Myronenko18 & $0.590 \pm 0.065$ & $0.758 \pm 0.045$ & $\textbf{0.774} \pm 0.057$ \\
\bottomrule
\end{tabular}
\label{table3}
\end{center}
\end{table}

Qualitative comparisons of ground truth surgical results from surgeons and generated results by different loss function in U-net on three bones are shown in Fig. \ref{fig:final}. All outputs generated by RPDL are graded the highest among all loss functions, evaluated by expert surgeons following the \emph{CM assessment scale}. Generated outputs by RPDL own (1) shapes that are more similar to those from surgeons; (2) depth of every surgical part that is closer to that from surgeons; (3) less deficiency than those by WCEL and DL. RPDL performs better than other loss functions measured by both evaluation metrics and experts. In addition, different models with same loss functions also perform diversely in Experiment 1. Myronenko18 performs the best overall, followed by U-net, Isensee17, and V-net. There may be several reasons for it. First, we use same architectures of all models in their original papers with minor change, which may not be the best for this task, respectively. Second, we do not do fine tuning of hyperparameters for every model in order to obtain their best performance. Instead, most hyperparameters in four models are the same. In consequence, models perform well in other tasks may not perform competitively here. For example, Isensee17 and V-net should not perform worse than U-net after delicate design of architectures and fine tuning of hyperparameters. However, the diverse performance of models with same loss functions is beyond the discussion of this paper. Our goal is to fairly compare the performance of RPDL with other loss functions in every representative model with same architectures and hyperparameters.

\subsection{Experiment 2: Comparison between Models and Surgeons}

The second set of experiments picks 9 surgeries performed by the same surgeon out as baselines at every time in the 7-fold cross validation. There are 6 segmentation annotation images used for training in Experiment 2. First, we evaluate the similarity of surgeries performed by surgeons in the testing set with those in the training set. The overall similarity among them is $\textbf{0.698} \pm 0.088$ measured by $Dice \pm std$ and $\textbf{0.780} \pm 0.069$ measured by $RPD \pm std$, which are regarded as the overall baseline provided by surgeons. We then train DCNN models with different loss functions to provide segmentation outputs similar to all surgeries in the training set. The goal is to evaluate if DCNN models can provide outputs more similar to segmentation annotation images than surgeons in the testing set. DCNN models are trained to summarize the commonality of all surgeons (6 surgeons) in the training set and provide the most similar segmentation output to them on every bone.

The overall results are consistent with  each  other  measured by both Dice and RPD coefficients (Table \ref{table4}). Both DL and RPDL performs better than the overall baseline of surgeons in the testing set while WCEL performs worse than it, evaluated by both evaluation metrics and expert surgeons. It again proves that WCEL is not capable of addressing the high-imbalance problem well. However, the performance of RPDL is very close to that of DL, with approximately $1\%$ overall difference, measured by either Dice, or RPD. It does not mean that DL is definitely better than RPDL in Experiment 2. Actually, RPDL enhances the ability of DCNN models to find the commonality among non-unique segmentation outputs and get rid of outside regions. Models with RPDL pays more attention to common regions of surgeries provided by more surgeons and pays less attention to those provided by fewer surgeons. They are even penalized heavily when models drill out of common regions of all surgeons. As a result, segmentation outputs provided by RPDL are slightly different from those provided by DL, which can be either larger or smaller. Experiment 2 is designed to evaluate if DCNN models can provide similar surgeries to those provided by different surgeons in the training set, with surgeons in the testing set as the overall baseline. It is not as challenging as Experiment 1 because DCNN models only need to provide segmentation outputs similar to all surgeons on same bones in the training set. They are not trained to perform surgeries on new bones in Experiment 2. Both DL and RPDL can achieve good performance on extracting the similarity of surgeries from different surgeons on the same bone. In conclusion, RPDL is able to enhance the ability of models to extract the similarity among non-unique segmentation outputs corresponding to same input images. Both DL and RPDL can extracting their similarity better than the overall baseline of new surgeons.

\begin{table}[tbp]
\scriptsize
\caption{Overall results evaluated on different surgeons, Experiment 2 (up: $Dice \pm std$; down: $RPD \pm std$).}
\begin{center}
\begin{tabular}{c|c|c|c|c}
\toprule
& Training loss & WCEL & DL & \textbf{RPDL} \\
\midrule
\multirow{4}{*}{Dice} & U-net & $0.618 \pm 0.109$ & $0.798 \pm 0.089$ & $0.781 \pm 0.104$ \\ 
& V-net & $0.639 \pm 0.114$ & $0.803 \pm 0.089$ & $0.782 \pm 0.105$ \\
& Isensee17 & $0.638 \pm 0.113$ & $0.801 \pm 0.087$ & $0.782 \pm 0.102$ \\
& Myronenko18 & $0.649 \pm 0.115$ & $0.761 \pm 0.077$ & $0.745 \pm 0.093$ \\
\midrule
\multirow{4}{*}{RPD} & U-net & $0.7 \pm 0.078$ & $0.884 \pm 0.068$ & $0.878 \pm 0.078$ \\ 
& V-net & $0.736 \pm 0.083$ & $0.888 \pm 0.068$ & $0.879 \pm 0.079$ \\
& Isensee17 & $0.735 \pm 0.082$ & $0.887 \pm 0.067$ & $0.879 \pm 0.077$  \\
& Myronenko18 & $0.755 \pm 0.084$ & $0.852 \pm 0.061$ & $0.845 \pm 0.072$ \\
\bottomrule
\end{tabular}
\label{table4}
\end{center}
\end{table}

\subsection{Advantages and Disadvantages}
There are some pros and cons of RPDL. The main advantage of RPDL is its excellent performance in the challenging scenario (Experiment 1). It outperforms WCEL and DL in all representative DCNN models (Table \ref{table3}). It is able to enhance the commonality of non-unique segmentation regions with limited data and large variety. Second, it does not increase the complexity of model architectures. RPDL is a new loss function, which can be used in any DCNN model. All DCNN models with RPDL can converge and outperform those with other loss functions with the same time in Experiment 1. However, there are also limitations of RPDL. First, the deviation of outputs provided by RPDL is larger than that provided by DL. Moreover, RPDL does not outperform DL in the easy scenario (see Table \ref{table4} in Experiment 2). The main reason is that RPDL tends to enhance the commonality of non-unique segmentation outputs and reward or penalize deviated regions provided by different surgeons.

\section{Conclusions}
In this paper, we define a new semantic segmentation problem, where one input image can correspond to multiple segmentation annotation outputs. We propose the reward-penalty Dice loss (RPDL) as the loss function for all deep convolutional neural networks (DCNN) in non-unique segmentation tasks. We demonstrate advantages of RPDL over existing loss functions on our collected cortical mastoidectomy (CM) dataset. Experimental results show that RPDL can outperform other loss functions dominantly in the challenging scenario and surpass surgeons in the easy scenario. We extend the potential of semantic segmentation in non-unique segmentation tasks. For future work, we plan to do research on (1) learning the reward-penalty map (RPMap) intelligently in RPDL and (2) the stability of training models with RPDL.

\section*{Acknowledgment}
We truly appreciated great cooperation with $7$ anonymous surgeons at the Royal Victorian Eye and Ear Hospital, who helped us perform, validate and grade CM surgeries in the VRTBS simulator. We were grateful for $9$ anonymous patients who provided their CT-scan temporal bone images for our research. This research was supported by the Melbourne Research Scholarship. This research was undertaken using the LIEF HPC-GPGPU Facility hosted at the University of Melbourne. This Facility was established with the assistance of LIEF Grant LE170100200.

\bibliographystyle{IEEEtran}
\bibliography{IEEEabrv,IJCNN19}

\end{document}